\documentclass{aa}
\usepackage[varg]{txfonts}
\usepackage{graphicx}

\bibpunct{(}{)}{;}{a}{}{,}
\usepackage{natbib,twoopt}
\usepackage[breaklinks=true]{hyperref} %% to avoid \citeads line fills
\bibpunct{(}{)}{;}{a}{}{,} %% natbib format for A&A and ApJ
\makeatletter
\newcommandtwoopt{\citeads}[3][][]{\href{http://adsabs.harvard.edu/abs/#3}%
{\def\hyper@linkstart##1##2{}%
\let\hyper@linkend\@empty\citealp[#1][#2]{#3}}}
\newcommandtwoopt{\citepads}[3][][]{\href{http://adsabs.harvard.edu/abs/#3}%
{\def\hyper@linkstart##1##2{}%
\let\hyper@linkend\@empty\citep[#1][#2]{#3}}}
\newcommandtwoopt{\citetads}[3][][]{\href{http://adsabs.harvard.edu/abs/#3}%
{\def\hyper@linkstart##1##2{}%
\let\hyper@linkend\@empty\citet[#1][#2]{#3}}}
\newcommandtwoopt{\citeyearads}[3][][]%
{\href{http://adsabs.harvard.edu/abs/#3}
{\def\hyper@linkstart##1##2{}%
\let\hyper@linkend\@empty\citeyear[#1][#2]{#3}}}
\makeatother

\begin{document}

\title{Magnetic flux structuring of the quiet Sun internetwork 
 }
 \subtitle{Center-to-limb analysis of solar-cycle variations}

  \author{
   M. Faurobert\inst{1}
   \and
   G. Ricort\inst{1}
  }

   \institute{ Universit\'{e} C\^{o}te d'Azur,  Observatoire de la C\^{o}te d'Azur, CNRS UMR 7293 J.L. Lagrange Laboratory,
   Campus Valrose, 06108 Nice, France\\
              \email{marianne.faurobert@oca.eu, gilbert.ricort@oca.eu}        
             }
                
\date{}
\titlerunning{Magnetic flux structuring in internetwork regions}
 \authorrunning{Faurobert \& Ricort}
 
\abstract
{The  small-scale magnetism of the quiet Sun has been investigated by various means in recent decades. It is now well established that 
the quiet Sun contains in total  more magnetic flux than active regions and represents an important reservoir of magnetic energy.  But the nature and evolution of these fields remain largely unknown.}
{ We investigate the solar-cycle and center-to-limb variations of magnetic-flux structures at small scales in internetwork regions of the quiet Sun.}
{We used Hinode SOT/SP data from the irradiance program between 2008 and 2016. Maps of the magnetic-flux density are derived from the center-of gravity method applied to the circular polarization profiles in the FeI 630.15 nm and FeI 630.25 nm lines.  To correct the maps from the  instrumental smearing of the telescope, we applied a deconvolution method based on a principal component analysis of the line profiles 
and on a Richardson-Lucy deconvolution of their coefficients. We took defocus effects and the diffraction of the SOT telescope into account.  We then performed a spectral analysis of the spatial fluctuations of the magnetic-flux density in 10'' x 10'' internetwork regions spanning a wide range of latitudes from $\pm 70^\circ$  to the equator.}
{  At low and mid latitudes  the power spectra normalized by the mean value of the unsigned flux in the regions do not vary significantly with the solar cycle. However at solar maximum for one scan in the activity belt showing an enhanced network,  a marginal increase in the power of the magnetic fluctuations is observed at granular and larger scales in the internetwork. At high latitudes, we observe variations  at granular and larger scales where the power decreases at solar maximum. At all the latitudes  the power of the magnetic fluctuations at scales smaller than 0.5''  remain constant throughout the solar cycle. 
 }
{ At the equator the unsigned flux density is related to the vertical component of the magnetic field, whereas at high latitudes this flux density is mainly related to the horizontal component and probe higher altitudes. Our results favor a small-scale dynamo that operates in the internetwork, but they show that the global dynamo also contributes to the internetwork fields. At solar maximum the high-latitude horizontal internetwork fields seem to be depleted from the structures at granular and larger scales that are seen at solar minimum, whereas the internetwork within enhanced network regions show more structures at those scales than 
at solar minimum. }


\maketitle

\section{Introduction}
\label{Sec:Intro}
The internetwork (IN) refers to the regions of the solar surface that are outside of active regions and the magnetic network. The IN is  pervaded by magnetic flux patches at various scales.  \citet{Gosic2014} estimate that 15\% to 20\% of the total quiet Sun flux is in the form of IN elements. The total flux content of the quiet Sun was shown to 
be slightly higher than the total flux of solar active regions at solar maximum \citep{Jin2011}. Therefore IN magnetic fields bring large amounts of magnetic flux to the solar surface and may play an important role  in the heating of the chromosphere. These magnetic fields have been extensively studied observationally and theoretically  in recent decades. A very thorough review of the numerous and sometimes contradictory observational investigations of the quiet Sun magnetism is presented in \citet{BellotRubio2019}. This review underlines the difficulty of observing these fields that vary on small scales in space and time and give rise to weak  polarimetric signals. But the gain in spatial resolution and polarization sensitivity in the measurements have allowed us to make significant progress, in particular thanks to space-based observations on the Hinode satellite and the SUNRISE balloon. New sophisticated spatially coupled inversion techniques have also recently been developed by \citet{Danilovic2016} and applied to spectropolarimetric observations performed at disk center on board Hinode. According to \citet{BellotRubio2019}, it is now possible to propose a unifying view of  observational studies. The IN magnetic fields  emerge at the solar surface on the form of magnetic loops with various sizes and lifetimes. The presence of many of these loops with an isotropic distribution of azimuth would agree with the probability distribution functions  (PDF) of the magnetic strength and orientation that are derived from spectropolarimetric inversions. The emergence of such loops has been observed by several authors as the transient appearance of a linear polarization patch, followed by two circular polarization patches (of opposite polarity) that separate from each other with time.  Hundreds of such events have been recorded by the Imaging Magnetograph eXperiment (IMaX) on SUNRISE. The simulations also show that magnetic loops are a consequence of the interaction of convective flows with the magnetic field. The simulation by \citet{Stein2011} shows a hierarchy of loop-like structures of smaller and smaller sizes as the loop apex approaches the solar
 photosphere.

However, as noted in \citet{BellotRubio2019}, the origin of  IN magnetic fields is still debated.  They may be due to the recycling of decaying active regions or generated by the solar dynamo in the deep convection zone, alternatively they may be produced by a local dynamo operating at the solar surface. Contradictory results have been obtained in different studies on this issue. \citet{Katsukawa2012} do not find indications of local dynamo action in their analysis of the velocity and magnetic power spectra from Hinode observations. Other works by \citet{Lites2014} and \citet{Buehler2013} reach the opposite conclusion. \citet{Buehler2013} find no variations of weak IN signals in Hinode data at disk center with the solar cycle; this is in favor of a local-dynamo origin. 
 \citet{Lites2014} study the full range of latitudes using synoptic maps taken by Hinode during the period 2008 - 2013. 
These authors find no variations at mid-latitudes for the weakest flux.  At higher latitudes, and especially near the poles, 
they detect clear changes in the magnetic flux associated with the solar cycle and the polarity reversal.

In this paper we address again the issue of the origin of IN magnetic fields. As in  \citet{Lites2014}  we investigate possible solar-cycle and latitudinal variations of the properties of IN magnetic fields. We also use synoptic data from the 
Hinode irradiance program between 2008 and 2016, but we implement a  different method. As the spatial structuring of the magnetic fields reflects the physical mechanisms at play, we analyze the Fourier power spectrum of the unsigned magnetic flux fluctuations on selected 10'' x 10'' regions of the quiet sun, away from the network patches.  We restrict our study to the longitudinal flux density that we derive from the intensity and circular polarization profiles in the Fe I 630 nm lines using the center-of-gravity method \citep[for extensive tests of the method, see][]{Uitenbroek2003}.  As  in \citet{Katsukawa2012}, we correct the data from the effect of the instrumental point spread function (PSF) of the SOT telescope. We also correct a time-varying defocus. Instead of correcting the magnetic-map power spectrum,  we apply the PSF correction to the observed  intensity and polarization profiles; the magnetic maps are not observed by the telescope. To do so we follow the techniques presented in \citet{QuinteroNoda2015}; that is, we perform a decomposition of the  Stokes profiles with a principal component analysis and we apply a Richardson-Lucy deconvolution to the maps of the coefficients of the decompositions.
This also allows an efficient filtering of the noise affecting the Stokes profiles. We then compute the magnetic flux density from the corrected Stokes profiles using the center-of-gravity method.
 
In the following section we explain how we obtain the corrected magnetic maps, the third section is devoted to the presentation of their Fourier power spectra, and the fourth section to the conclusions of this work.

\section{Maps of the longitudinal magnetic flux density}
The two-dimensional data are obtained by scanning the slit of the spectrograph on the surface of the Sun and  recording the Stokes profiles (I, Q, U, V) on each point along the slit for a set of discrete wavelength points around the FeI 630 nm lines with a spectral sampling of 2.15 pm pixel$^{-1}$ . 

\subsection{Hinode data and estimate of the defocus}
We use center-to-limb
observations that are part of the irradiance survey program HOP 79  of  Hinode/SP, in which pole-to-pole scans were performed with a pixel size of  0.32" /pixel along the spectrograph slit and  0.1476'' steps in the east-west direction. The pixel size along the slit results from an on-board rebinning of two 0.16'' pixels. To study the long-term evolution of IN
magnetic fields, we chose to analyze one set of observations per year between 2008 and 2016. We thus follow more than half a cycle, starting at the minimum of cycle 24,  in 2008 and going beyond its maximum that took place in 2014.  We used level-1  Hinode/SP data  that are available on the HAO database\footnote{ 10.5065/D6T151QF.}; the data preparation package is described in \citet{LitesSP-PREP2013}. In this work we use the runs of October  2008,   July 2009,   July 2010,  August 2011, July  2012, July  2013,  May 2014,  August 2015, and  July 2016.

Before starting the data analysis we first interpolated the Stokes profiles  to a pixel size of 0.16'', that is, the pixel-size of the SOT/SP camera, in the $x$ and $y$ directions  by a simple bilinear interpolation. This is because in this work we then compute 2D Fourier transforms of images that require square pixels.  
 We checked that this interpolation has a very small effect on the average and on the standard deviation of the intensity and Stokes V signals on the scanned regions.

A time-varying defocus of the SOT telescope has been reported by various authors \citep[see, e.g.,][]{Lites2014}. In this work we estimate the defocus as detailed in \citet{Faurobert2015}.  A defocus modifies the modulation transfer function (MTF) of the instrument,
which affects the Fourier spectra through a multiplicative function that can be  easily modeled  as  a varying phase term over the entrance pupil of the telescope. To estimate this phase term, we examined the Fourier power spectrum of the granulation in the continuum at 630 nm observed at the center of the solar disk and we compared it to a reference spectrum. In case of a defocus the ratio of both spectra shows a characteristic shape that allows us to estimate the corresponding phase term on the perturbed MTF.  As a reference, we chose the spectrum observed in 2015 that was the best in-focus data set. We checked the spectra ratios for the characteristic shape due to a defocus and we corrected the MTF of the other years by multiplying with the corresponding phase term.

\subsection{Principal component analysis of the Stokes profiles and deconvolution} 
Contrary to the typical procedure in image deconvolution, \citet{QuinteroNoda2015} proposed to work on the spectral dimension and not on the spatial dimensions. They assumed that the  Stokes profiles at each pixel can be written as a linear combination of an orthonormal  basis formed by eigenfunctions ${\Phi_i(\lambda)}$,
\begin{equation}
 S_u(\lambda,x,y) = \sum_{i=1}^{N_\lambda}\omega_i (x,y)\Phi_i (\lambda),
 \label{eq:coeff}
 \end{equation}
where $N_\lambda$ is the number of wavelength points along the spectral dimension. Usually only a few elements of the eigenfunctions are enough to reproduce the  Stokes profiles, thus we may  truncate the previous sum and only take the first $n_\lambda$ terms into account. Therefore, the  data may be described by a set of images $\omega_i$, which are built by projecting the Stokes profiles of each pixel on the basis functions. Then, assuming that the PSF does not vary with the wavelength, the observed Stokes profiles are written as
\begin{equation}
S_p(\lambda,x,y) = \sum_{i=1}^{n_\lambda}(\omega_i (x,y)* PSF(x,y))\Phi_i (\lambda) + N(x,y),
 \end{equation}
 where the star operator denotes the convolution and  $N(x,y)$ is some additive noise. The scalar product of the observed Stokes profiles with any of the eigenfunctions gives 
\begin{equation}
< S_p(\lambda,x,y), \Phi_k (\lambda)>  = \omega_k(x,y)* PSF(x,y) + N(x,y).
 \end{equation}
Therefore the maps of the coefficients $\omega_i(x,y)$ obtained from the observed Stokes profiles are affected by the same convolution with the PSF as the profiles. As a consequence we may recover the  coefficients $\omega_i$ by applying a PSF deconvolution to the observed maps, the  Stokes profiles are then reconstructed using Eq. (\ref{eq:coeff}). As in \citet{QuinteroNoda2015} we performed the deconvolution using the Richardson-Lucy algorithm and we derived the orthonormal basis  ${\Phi_i(\lambda)}$ from a principal component analysis. This choice is well suited for our problem because a limited number of  principal components are sufficient to account for the variability of the observed profiles. We found that eight components give a very good representation of the Stokes profiles, while allowing us to filter the noise that is mainly projected on less significant principal components. 

\subsection{Unsigned longitudinal apparent flux density} For our long-term study we chose to deal with the Stokes I and V profiles only. The reason is that the linear polarization profiles in the irradiance data of Hinode are much more noisy and  their quality is degrading after 2013. Furthermore we do not intend to use any sophisticated inversion method to recover the magnetic field components from the Stokes profiles. We prefer to implement the simple and robust center-of-gravity method, which was first introduced by \citet{Semel1970} and later extensively tested by \citet{Uitenbroek2003} by comparison with the magnetic flux in 3D magnetohydrodynamical simulations of the magneto-convection in the solar photosphere. 

In the center-of-gravity method the line-of-sight component of the magnetic field is derived from the wavelengths $\lambda_+$ and $ \lambda_-$ of the centroid of the right- and left-circularly polarized line components $I\pm V$. They are defined as
\begin{equation}
\lambda_{\pm} ={\int\lambda (I_{cont}-(I\pm V))d\lambda \over \int (I_{cont}-(I\pm V))d\lambda},
\end{equation}
where $I_{cont}$ denotes the continuum intensity. The longitudinal magnetic component is then obtained from the relation
\begin{equation}
B_{LOS}={\lambda_+ -\lambda_- \over 2}{4\pi m c \over e g_L\lambda_0^2},
\label{eq:blos}
\end{equation}
where $g_L$ is the line effective Land\' e factor and $m$ and $e$ are the mass and electric charge of the electron 
(in MKSA units), respectively.
When the magnetic  structure is not resolved by the instrument, Eq. (\ref{eq:blos}) gives the longitudinal apparent flux density, that is, the magnetic flux density averaged over the pixel area. We remark that because the wavelengths $\lambda_{\pm}$ are obtained from a ratio of observed intensities, they should not be greatly affected by a possible aging of the detectors or calibration issues. 

\begin{figure*}[ht]
\includegraphics[width=0.95\textwidth]{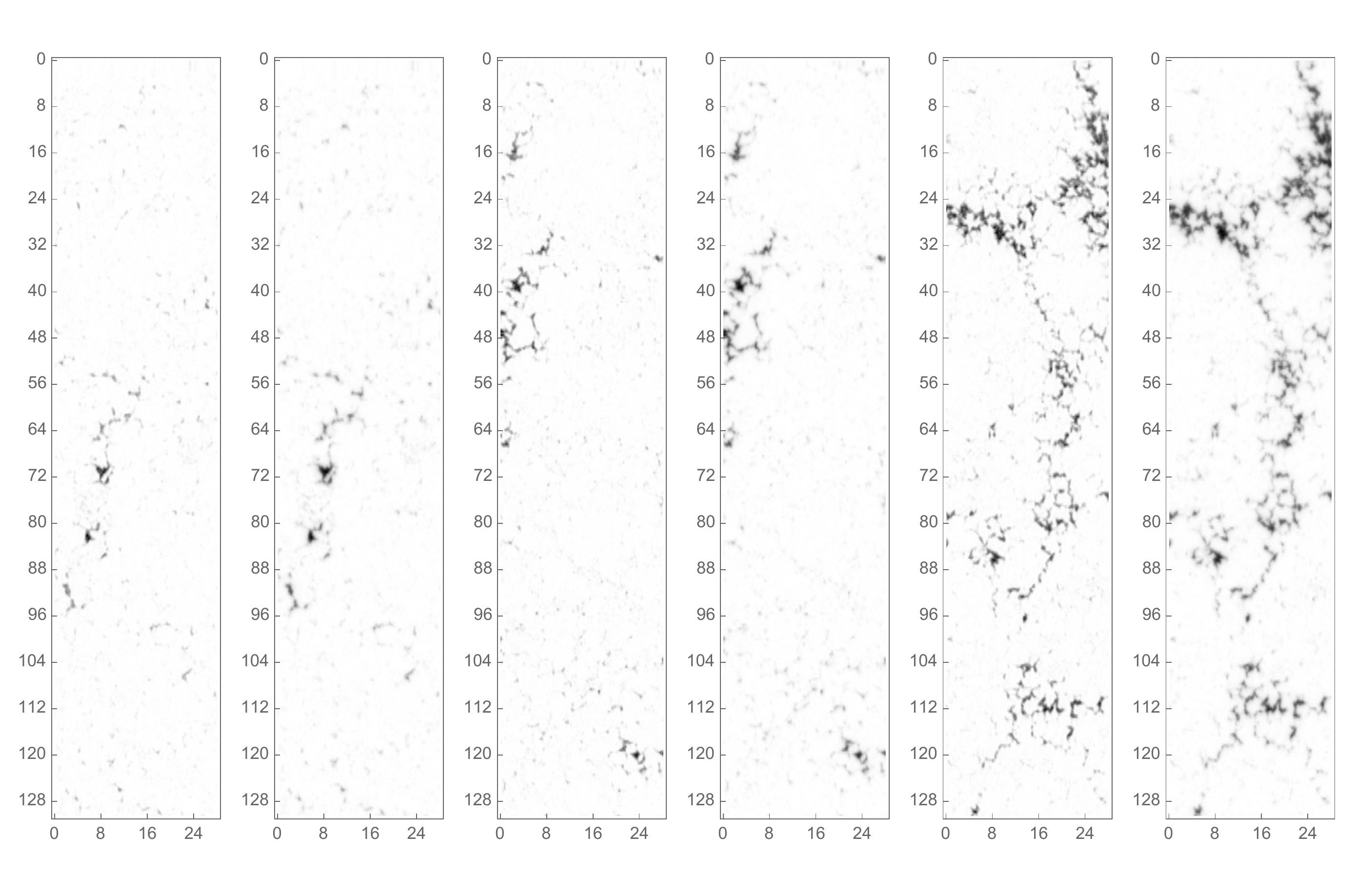}

\caption{Maps of the unsigned longitudinal magnetic flux density. From the left to the right: In 2008 at the center of the solar disk with correction for the PSF; the same without PSF correction  in 2014 at the center of the solar disk with PSF correction; the same  without PSF correction in 2014 at the latitude $\theta =-18^\circ$, showing enhanced network corrected for the PSF; and the same without PSF correction. The scales are in seconds of arc, the vertical axis is along the north-south polar axis, the horizontal axis is along the west-east direction.}
 \label{fig1}
\end{figure*}

Figure \ref{fig1} shows examples of uncorrected and de-convolved  unsigned magnetic flux maps obtained  
 in 2008 and in 2014. The sharpening effect of the deconvolution is quite visible.
 
\section{Power spectra of the magnetic maps}
As explained in the Introduction, we now focus on the magnetic structures at small scales in the IN. To avoid network patches we first select (10'' x 10'') regions located around unsigned magnetic flux local minima in the PSF-corrected maps. We chose to use one run of the irradiance program HOP79 per year from 2008 to 2016. Much more data are available because HOP 79 was performed almost once per month during the period, but we estimated that the long-term evolution of IN magnetic structures may be captured with a one year time step. More data will be analyzed in the future. For each year of our sample 
we have 20 maps such as those shown in Fig. \ref{fig1} located at various positions along the polar axis from the north to the south pole.
 We note that  because we used observations taken along the polar axis there is an ambiguity between center-to-limb variations  and latitudinal variations. But as shown in the following, the use of the two FeI lines at 630.15 nm and 630.25 nm that are formed at different altitudes provides us with some information on the height-dependence of the magnetic structures.  

\begin{figure}[ht]
\includegraphics[width=0.45\textwidth]{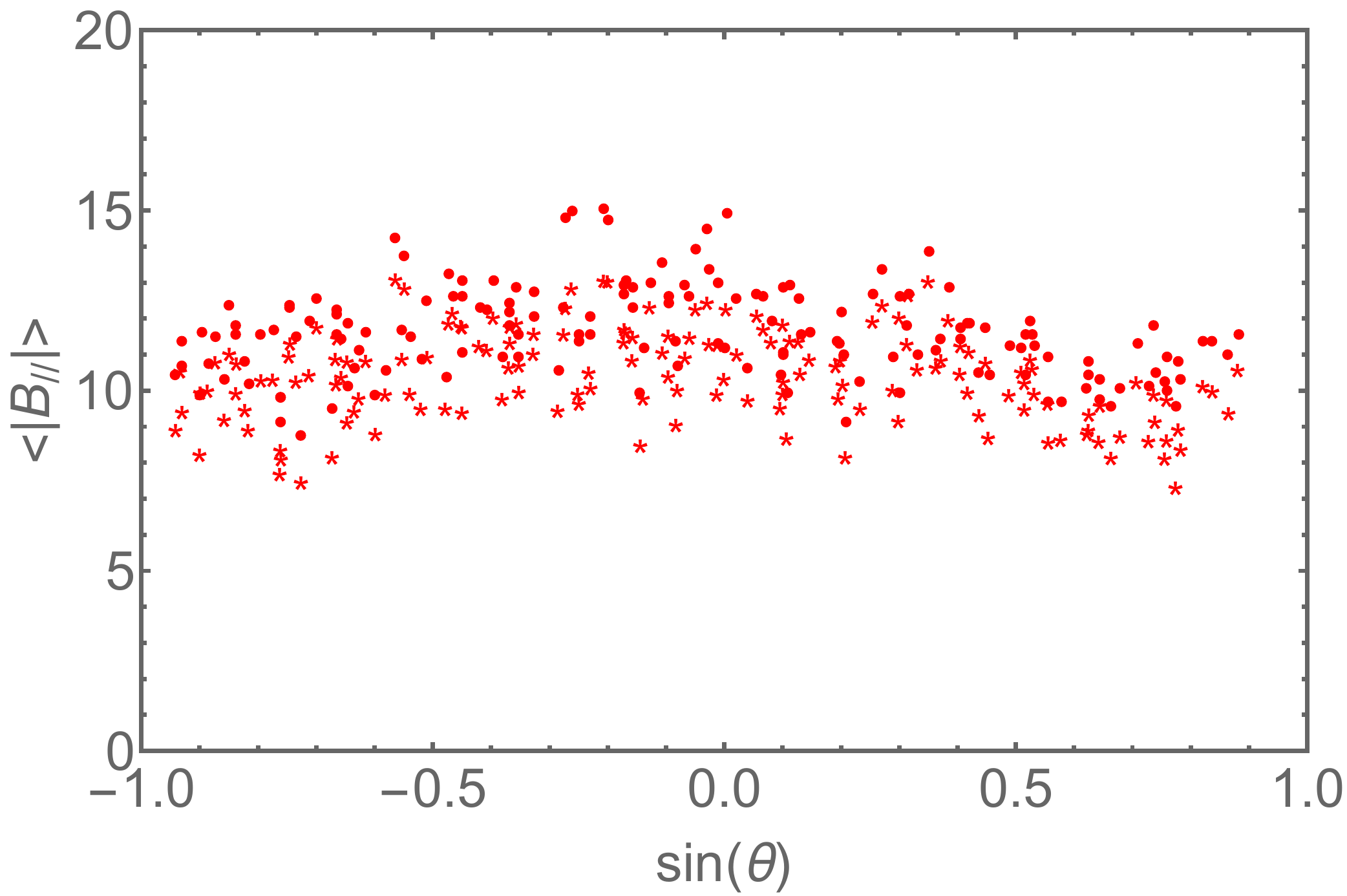}
\includegraphics[width=0.45\textwidth]{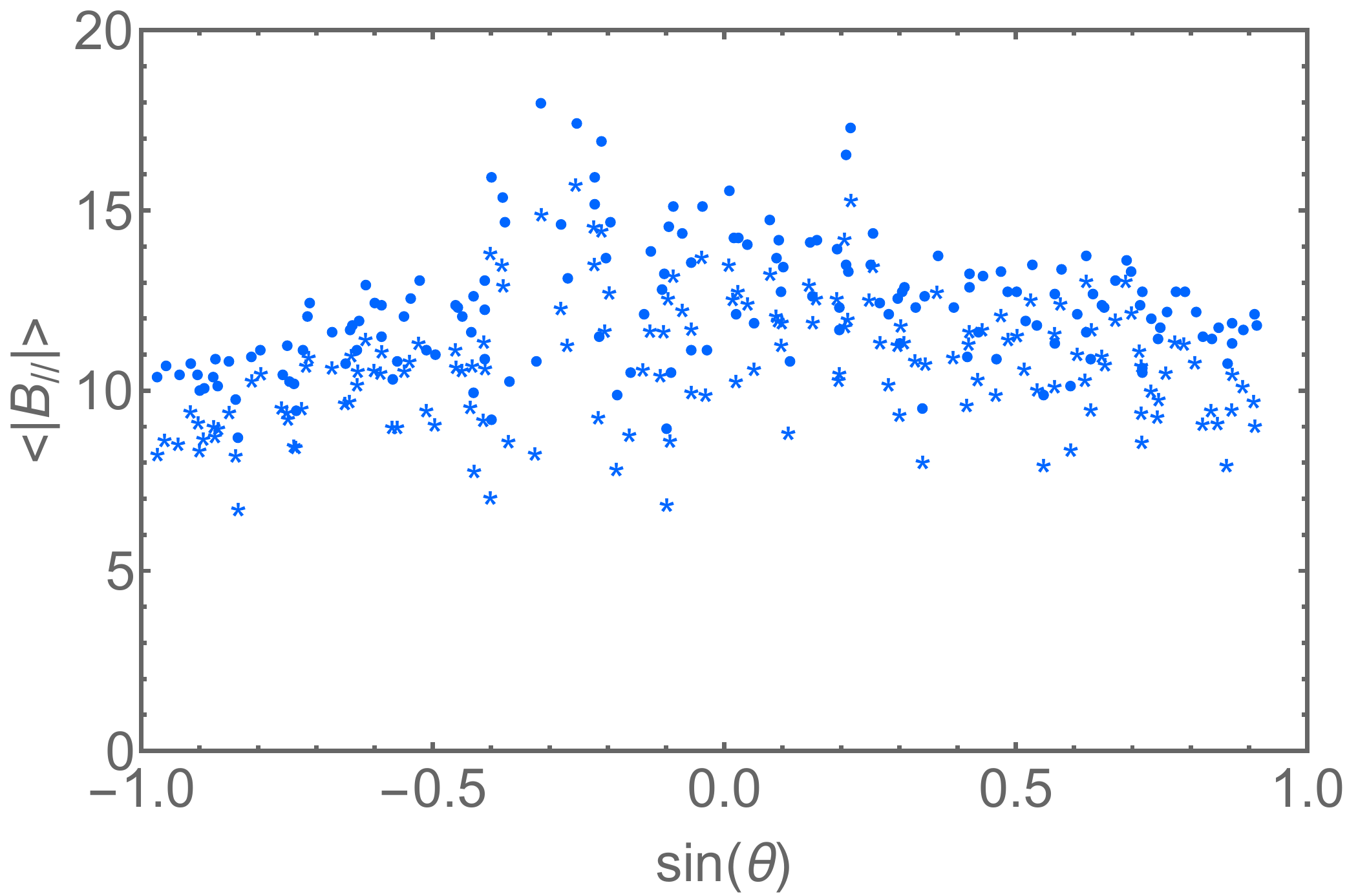}
 \caption{Mean value of the unsigned longitudinal magnetic flux density (in Mx cm$^{-2}$) in the selected IN regions vs. the sinus of their latitude. Top panel: The 2008 data set; bottom panel: The 2014 data set. The stars refer to the values obtained with the FeI 630.25 nm line and the dots to those obtained with the FeI 630.15 nm line}
  \label{fig2}
\end{figure}

\begin{figure}[ht]
\includegraphics[width=0.45\textwidth]{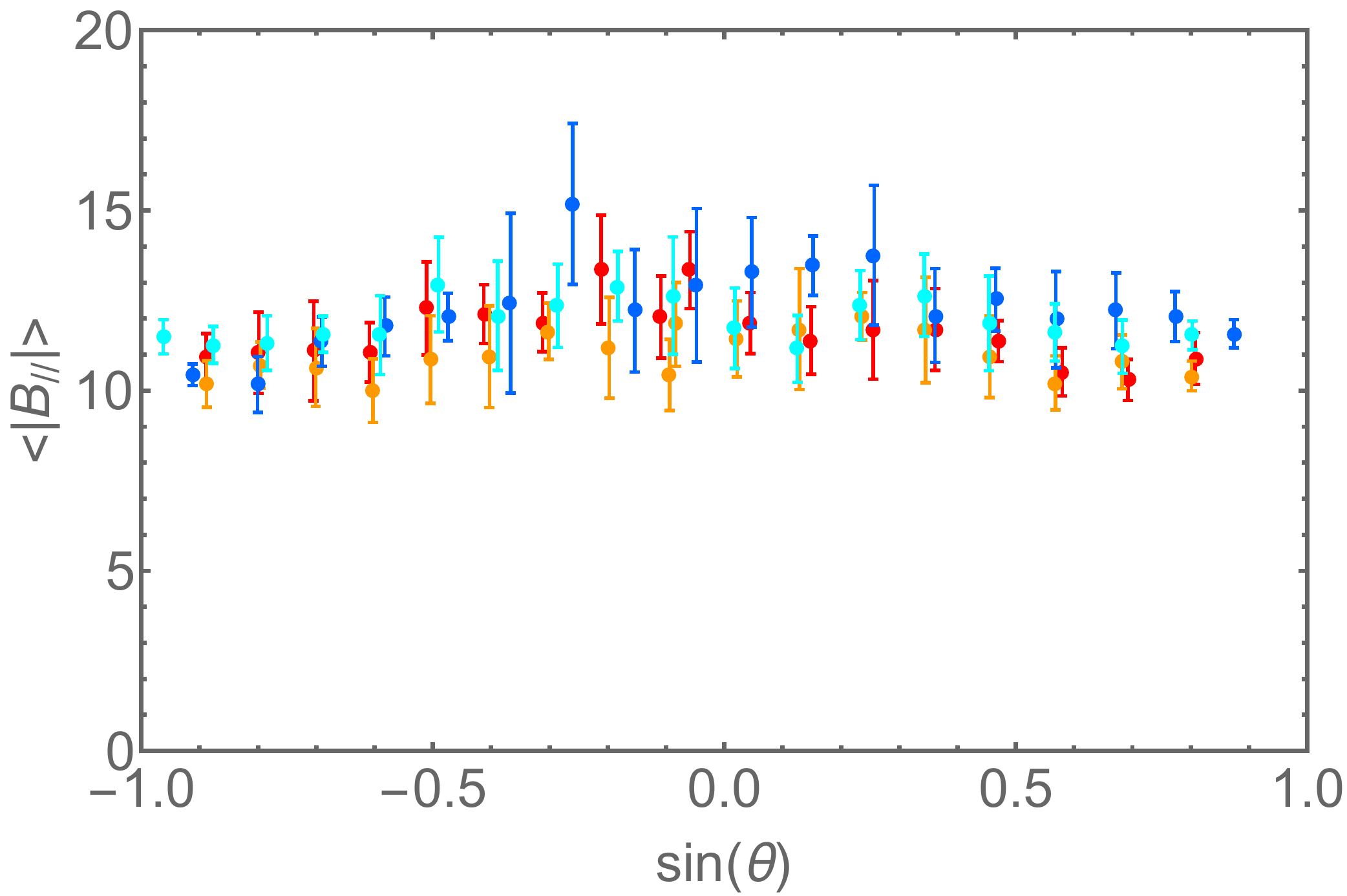}
 \caption{Average values of the mean unsigned longitudinal magnetic flux density (in Mx cm$^{-2}$)  on the nine IN regions selected in the magnetic maps  vs.s the sinus of the latitude of the map center.   Red dots: 2008; orange dots: 2009; light blue: 2013; and blue: 2014. The bars show one standard deviation intervals. }
  \label{fig2bis}
\end{figure}

\begin{figure}[ht]
\includegraphics[width=0.45\textwidth]{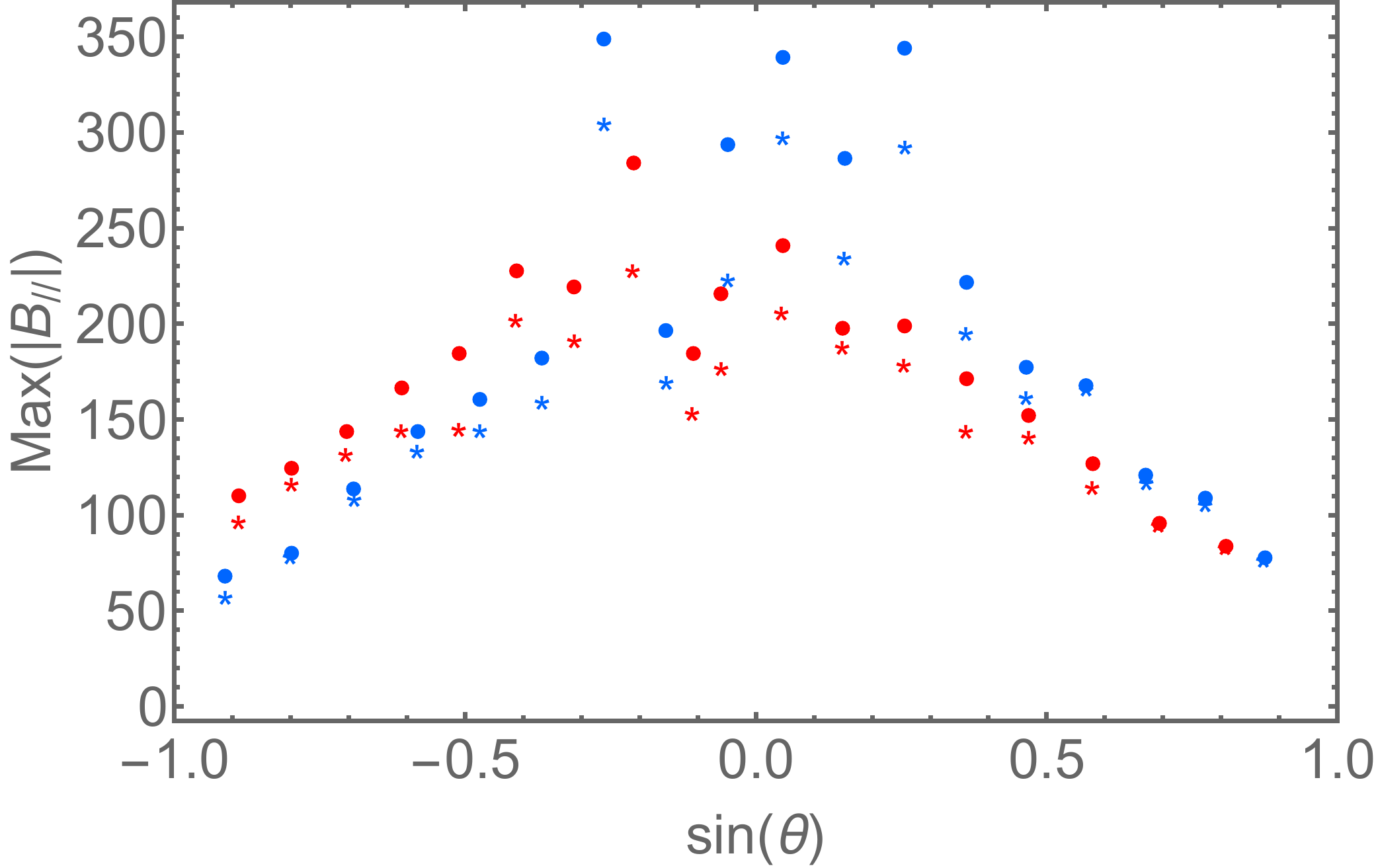}
 \caption{Maximum value of the mean unsigned longitudinal magnetic flux density (in Mx cm$^{-2}$)  in the IN regions in 2008  (red symbols) and 2014 (blue symbols) vs. the sinus of their latitude.  The stars refer to the values obtained with the FeI 630.25 nm line and the dots to those obtained with the FeI 630.15 nm line}
  \label{fig3}
\end{figure}

We selected nine IN regions on each of these maps. Figure \ref{fig2} shows the mean value of the unsigned flux density  on the selected regions as a function of the sinus of their latitude for the 2008 and 2014 data sets. The values are between 7 Mx cm$^{-2}$ and 15  Mx cm$^{-2}$ in the 2008 data and between 7 Mx cm$^{-2}$ and 18  Mx cm$^{-2}$ in 2014 at solar maximum. Figure \ref{fig2bis} shows the average of the mean unsigned flux  on the nine  IN regions that we selected on each magnetic map in four data sets obtained in 2008, 2009, 2013, and 2014, that is, during the minimum and the maximum of the last solar cycle.  The error bars show the interval at $\pm \sigma$ estimated from the dispersion of the values on the nine regions. We do not observe significant latitudinal variations of the average unsigned flux density in the IN regions, but the dispersion is slightly higher at solar maximum and the values are higher for the IN regions of one scan located at latitudes around -18$^\circ$, we return to this later in this paper. The mean value of unsigned flux in the selected regions is quite uniform around 10 $\pm 2$  Mx cm$^{-2}$ but some pixels have a much higher flux (on the order of several hundreds of  Mx cm$^{-2}$) than the average over the region. Figure \ref{fig3} shows the maximum values of the unsigned flux in the IN regions as a function of the sinus of the latitude in the 2008 and 2014 data. We clearly observe an increase at disk center; it is even more prominent at solar maximum. The increase at disk center  could be explained by the fact that the strongest fields are preferentially vertical, as already observed by several authors \citep[see the review by][]{BellotRubio2019} and the increase at solar maximum to the fact that they are partly due to decaying active regions.
The values measured with the FeI 630.25 nm line seem to be systematically smaller than those obtained with the FeI 630.15 nm line. This is also true for the mean values over the IN regions. We return to this later in this work.

In order to learn more about the physical origin of the IN magnetic flux we now focus on its spatial structures. A good tool to gain insight into spatial structures is to study  their Fourier power spectrum. We computed the 2D Fourier spectra of the IN unsigned flux density in the following way. We divided each selected region in four subregions of  32 px  x 32 px  (5'' x 5'') and computed the 2D Fourier transform and power spectrum on the four subregions. Then we  averaged  the power spectra obtained on the nine regions selected on each map. Thus the power spectra are obtained by averaging over 36 (5'' x 5'') IN regions for each of the 18 maps obtained  every year between 2008 and 2016; we disregarded the maps obtained too close to the poles because the spectrograph slit was intercepting the limb so they are smaller in size. Then we performed radial averages of the 2D power spectra as follows:
\begin{equation}
   { \cal E}( k )= 2\pi k \int_0^{2\pi} {\cal S}( k \cos\alpha, k \sin\alpha) {d\alpha\over 2\pi} ,   
    \label{Spectrad}
\end{equation}
where the spatial wavenumber $k= 2\pi\sqrt{u^2+v^2}$  is related to the spatial wavelength $\lambda_s$ of the fluctuations by $k=2\pi/\lambda_s$, and $u$ and $v$ denote the conjugate variables in the Fourier space of the $x$ and $y$ coordinates.
We estimated the statistical fluctuations of the power spectra from the dispersion observed for the nine IN regions on each maps.

We note that the radial average of the 2D power spectrum provides meaningful information about the spatial structuring  only when
the signal is isotropic on average (in the sense of the ensemble average). For solar images observed away from disk center, we have to account for the projection effect
leading to foreshortening of the images in the radial direction. We took this effect into account by averaging the 2D power spectra over
elliptical bands instead of circular bands.

We now discuss the spectral resolution of the power spectra. Working on regions of 5'' x 5'' amounts to multiplying the full maps by a square window of 5 '' x 5''; as a consequence the Fourier transform is convolved by a sinus cardinal function of width one-fifth arc-second$^{-1}$ on each dimension. So the resolution of the spectrum is reduced. The equivalent phenomenon is well known for time-dependent signals such as helioseismic signals. A long uninterrupted observing time is necessary to detect fine details in the power spectrum. 
In our case we chose to increase the signal-to-noise ratio on the spectra by a factor 2 by using 5'' x 5'' instead of 10'' x 10'' subregions at the price of loosing resolution on the spectra.  We do not expect to detect fine spectral features,  but we can nevertheless explore possible center-to-limb or long-term variations. The pixel size in the spectral domain is 0.195  arc-second $^{-1}$ (1/(32 x 0.16))), the largest accessible spatial frequency is 3.125 arc-second $^{-1}$ and the smallest is 0.195 arc-second $^{-1}$.

\begin{figure}[ht]
\includegraphics[width=0.45\textwidth]{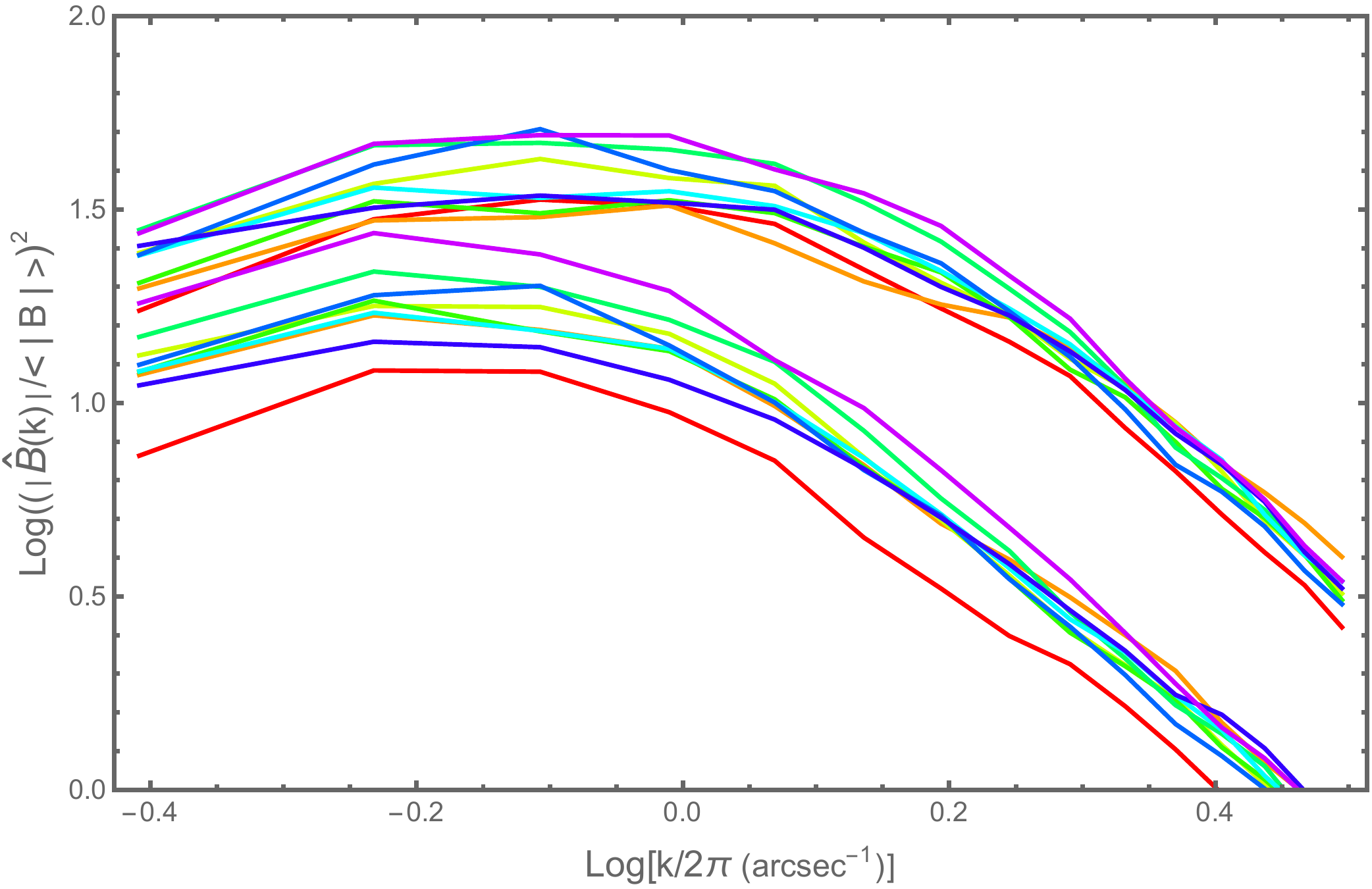}
 \caption{Power spectra of the spatial fluctuations of the unsigned magnetic flux density  in the IN regions at the center of the solar disk, for observations performed between 2008 and 2016 on a log-log scale.  The upper set of curves are obtained from PSF corrected maps, whereas the lower set is from uncorrected maps. Red line: 2008; orange: 2009; light green: 2010; green: 2011; deep green: 2012; light blue: 2013; blue: 2014; deep blue  line: 2015; and purple: 2016.}
  \label{fig4}
\end{figure}
\begin{figure}[ht]
\includegraphics[width=0.45\textwidth]{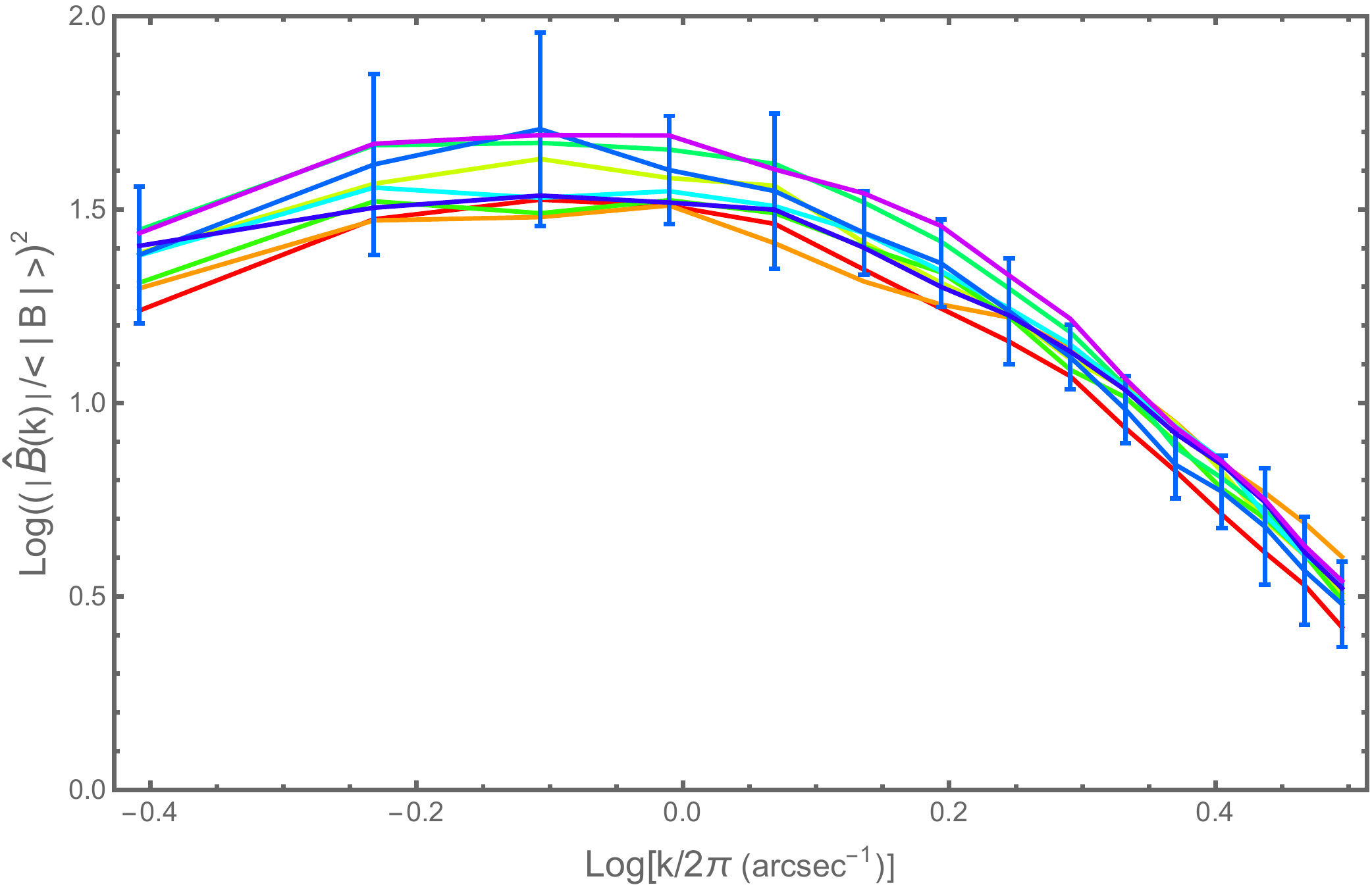}
\includegraphics[width=0.45\textwidth]{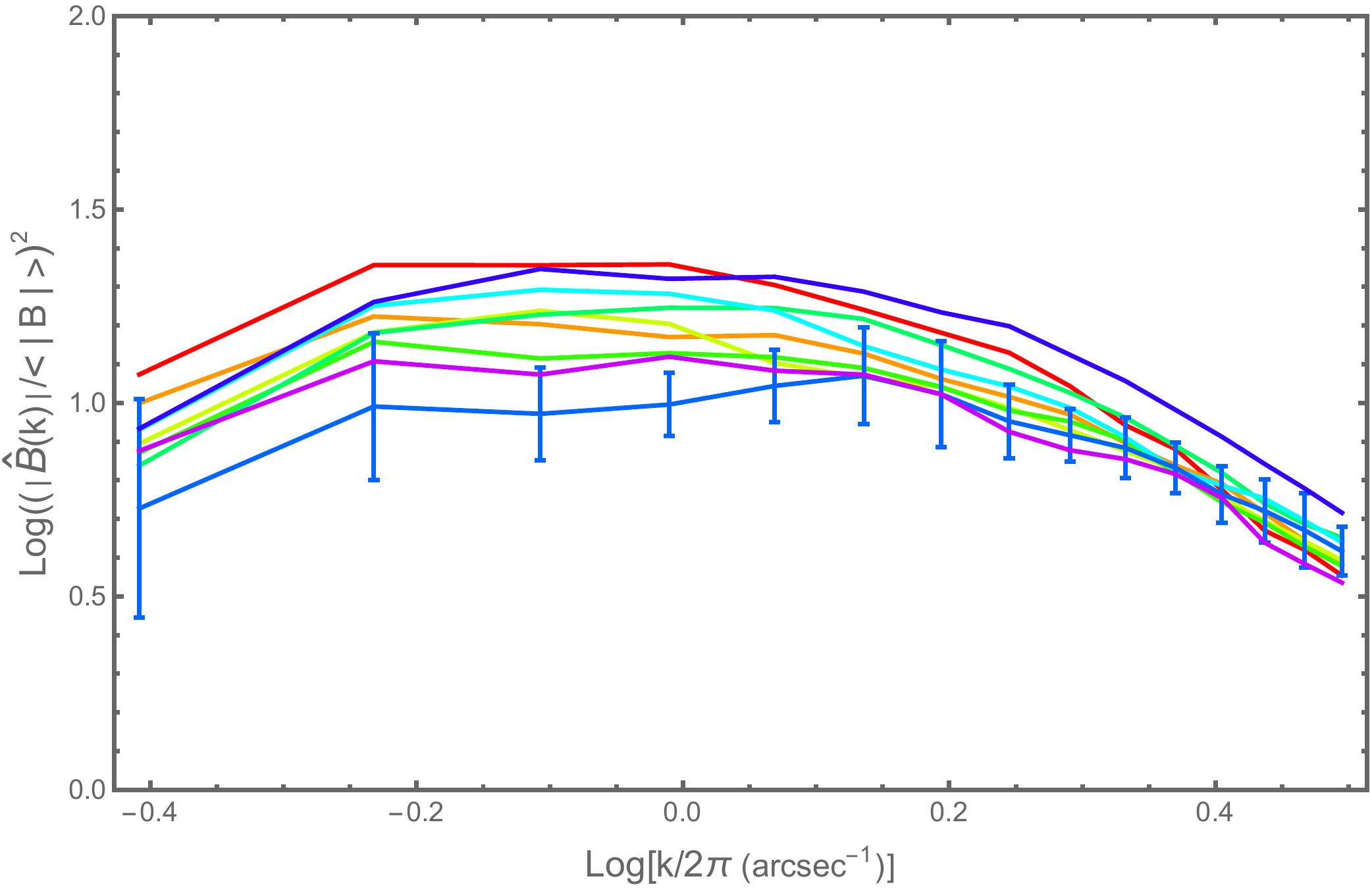}
\includegraphics[width=0.45\textwidth]{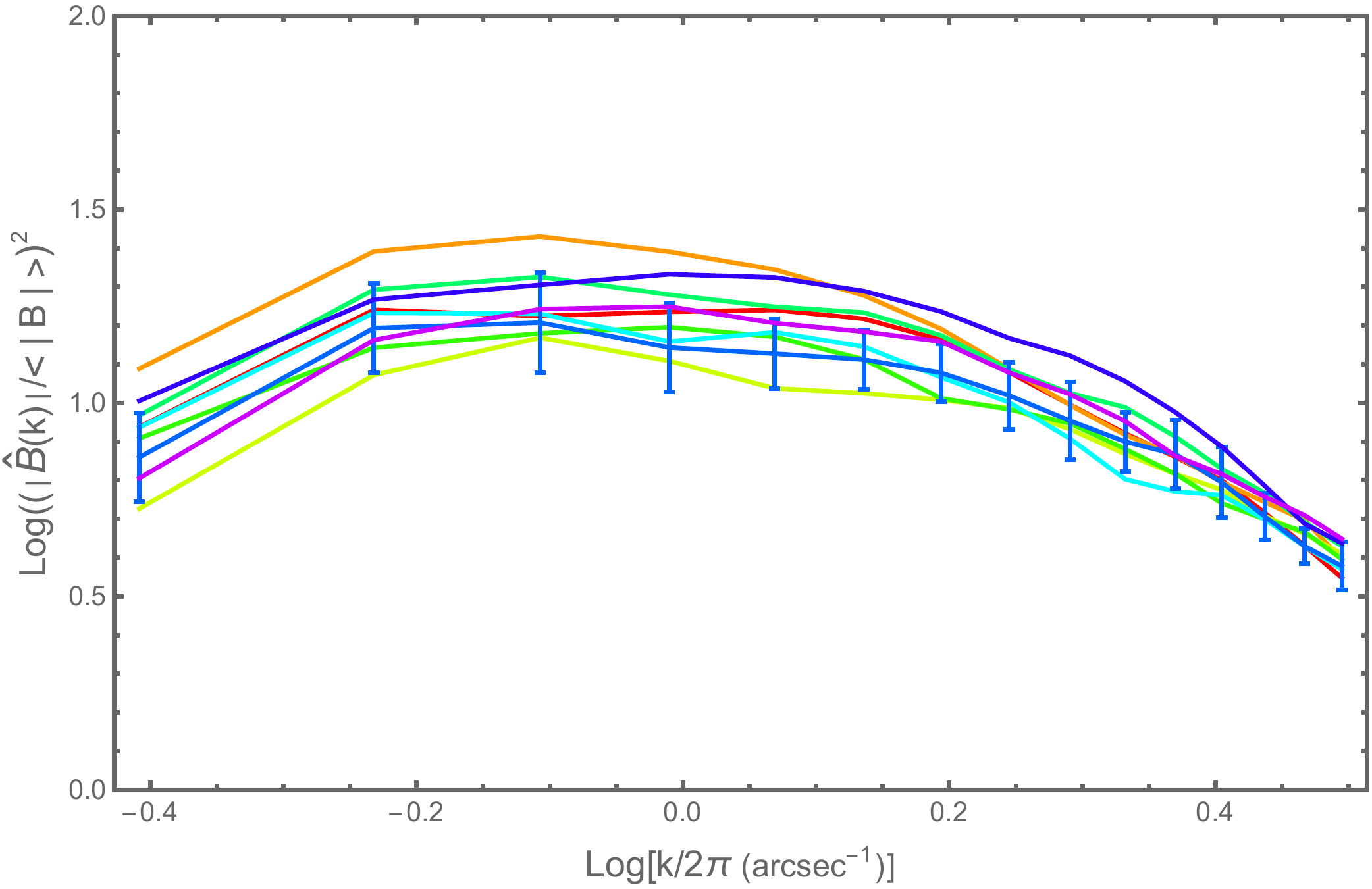}

 \caption{Power spectra of the spatial fluctuations of the unsigned magnetic flux density  in the IN regions for observations performed between 
 2008 and 2016,  on a log-log scale.  The bars show the $\pm\sigma$ intervals on the 2014 data set. Color coded as in Fig. \ref{fig4}. Top panel: At disk center; middle panel : at high latitude in the southern hemisphere; and bottom panel: at high latitude in the northern hemisphere.}
  \label{fig5}
\end{figure}

 
We now show the reduced power spectra, that is, the spectra divided by the square of the average unsigned flux density on the region. The integral of these curves would give us the equivalent of a squared contrast of the unsigned magnetic flux fluctuations. Figure \ref{fig4} shows the comparison of the power spectra computed on uncorrected maps and corrected ones. We clearly see the effect of the correction by the PSF; the power of the magnetic fluctuation is significantly increased when a deconvolution is applied to the  maps, as expected. The slope of the spectra at high spatial frequencies is modified too; the decrease of the power at high frequency is steeper and the position of the maximum is displaced to higher spatial frequencies. The power spectra of corrected maps show a broad maximum at sub-granular scales of about 900 km. Even though there seems to be a slight increase in the power of the fluctuations at solar maximum, this is not statistically significant when we take the standard deviation of the power spectra between the different IN regions at a given time into account (the error bars are shown in Fig. \ref{fig5}). We do not detect significant variations of the small-scale structuring of the longitudinal magnetic flux in the IN  at disk center between 2008 and 2016.

The reduced spectra obtained at disk center and close to the southern and northern poles at a latitude of $70 ^\circ$ are presented in Fig. \ref{fig5} , together with the error bars at $\pm \sigma$, where $\sigma$ denotes the standard deviation estimated on the nine IN regions selected on the maps.
At high latitudes we observe significant time variations of the power spectrum in opposition of phase with the global solar cycle, except at the largest spatial frequencies where the power spectra remain constant. 

\begin{figure}[ht]
\includegraphics[width=0.45\textwidth]{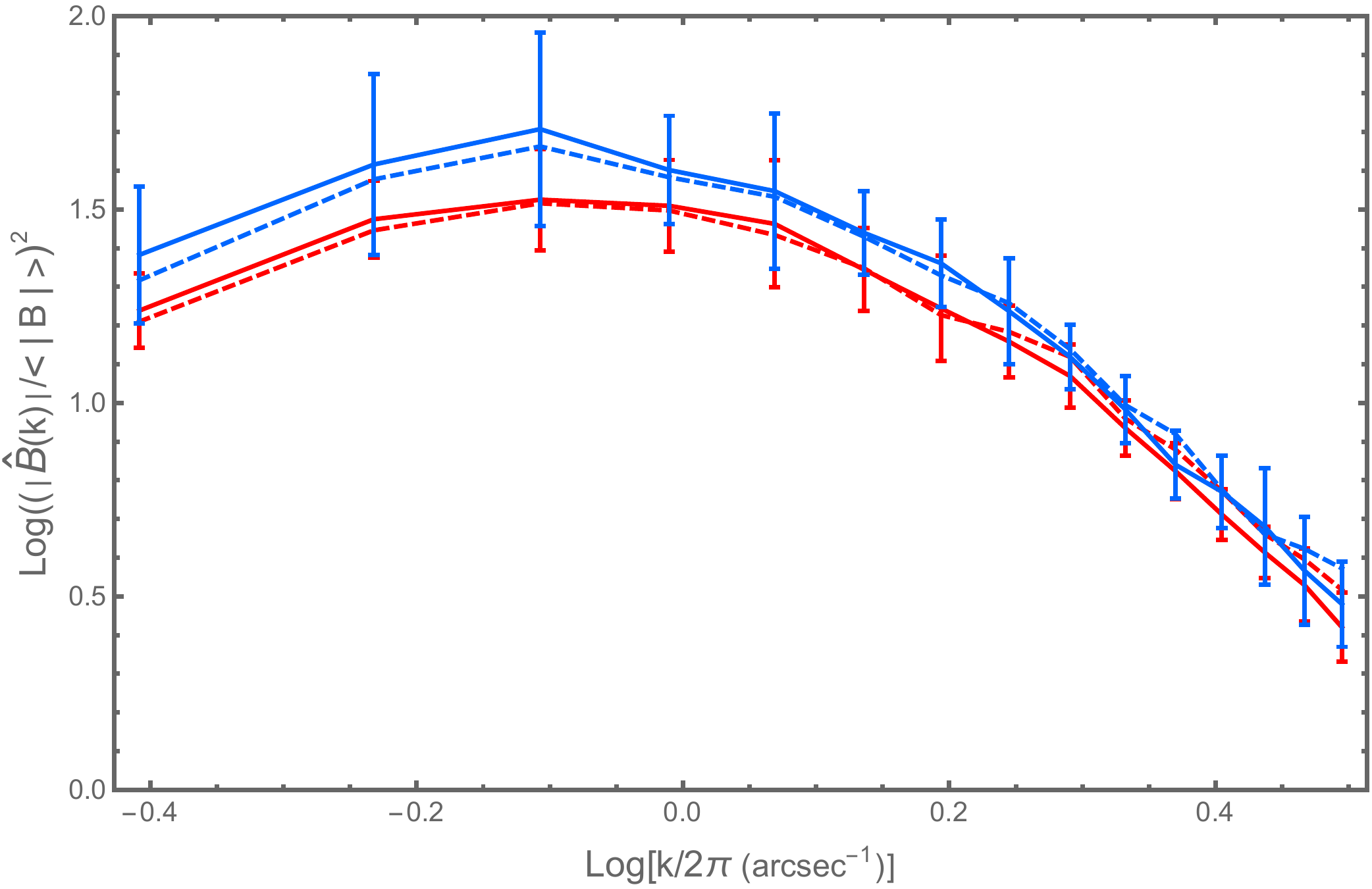}
\includegraphics[width=0.45\textwidth]{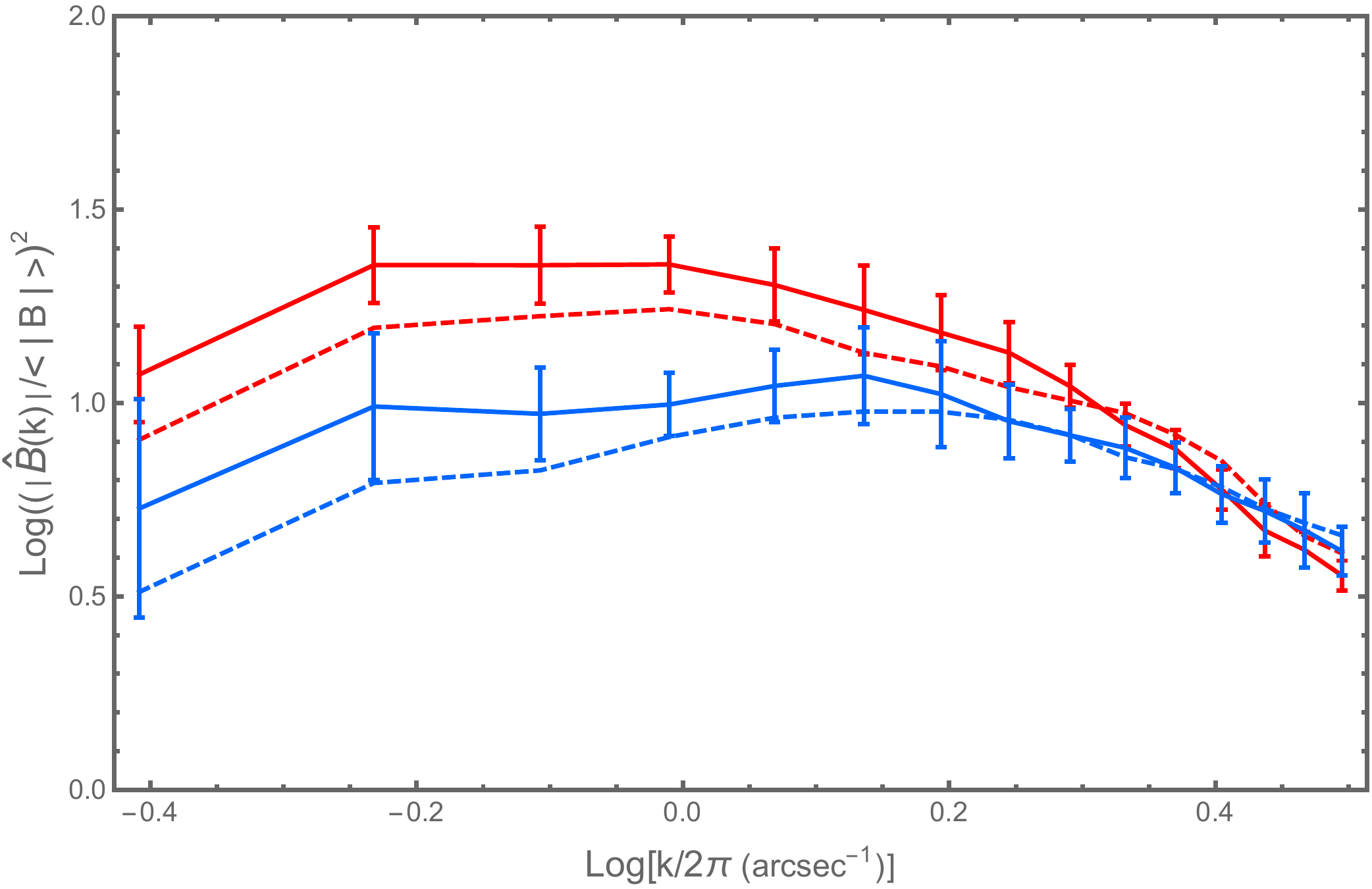}
\includegraphics[width=0.45\textwidth]{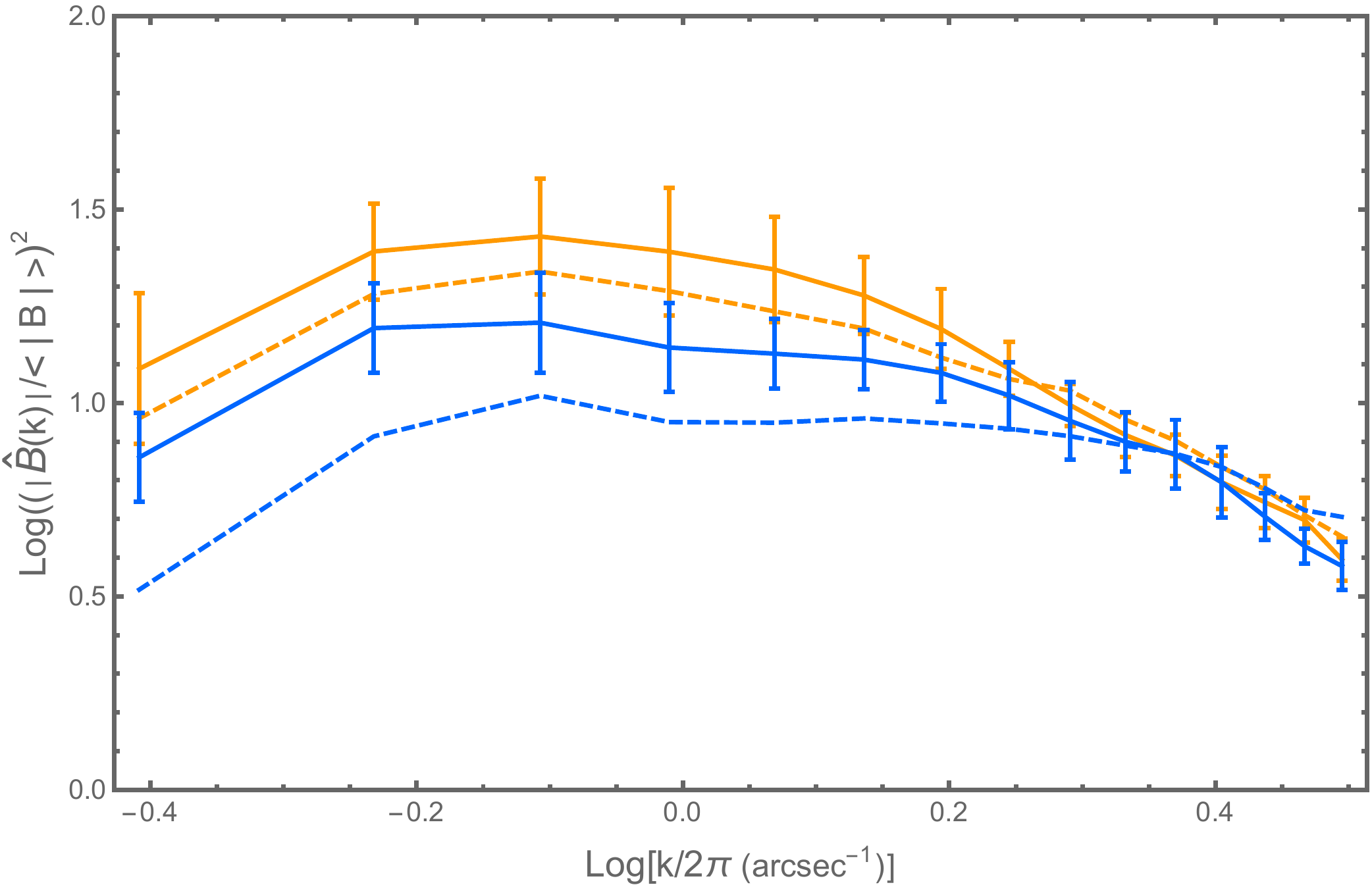}

 \caption{Power spectra of the spatial fluctuations of the unsigned magnetic flux density  in the IN regions, on a log-log scale. Top panel: at the center of the disk; middle panel: at high latitude in the southern hemisphere; and bottom panel: at high latitude in the northern hemisphere. The bars show the $\pm\sigma$ intervals. Blue lines: in 2014; red lines: in 2008; and orange lines: in 2009. Dashed (full) lines: from the FeI 630.15 nm (FeI 630.25 nm ) line.}
  \label{fig6}
\end{figure}

To better see the differences between the power spectra at solar minimum and maximum we present in Fig. \ref{fig6} the comparison between the reduced spectra obtained at the center of the disk and at high latitudes in the southern and northern hemispheres. 
We note that the northern hemisphere shows a maximum in the power of the magnetic flux fluctuations one year after the southern hemisphere, that is,  in 2009 instead of 2008. We also compare the spectra computed with the two FeI lines at 630.15 nm and 630.25 nm. We observed that the power of the fluctuation is higher on the maps obtained with the FeI 630.25 nm line than with the 630.15 nm line;  this may appear  weakly significant as compared to the statistical fluctuations of the spectra from one region to the other, but we observe this effect on all the  selected (10'' x 10'') IN regions and at all the latitudes.

\begin{figure}[ht]
\includegraphics[width=0.45\textwidth]{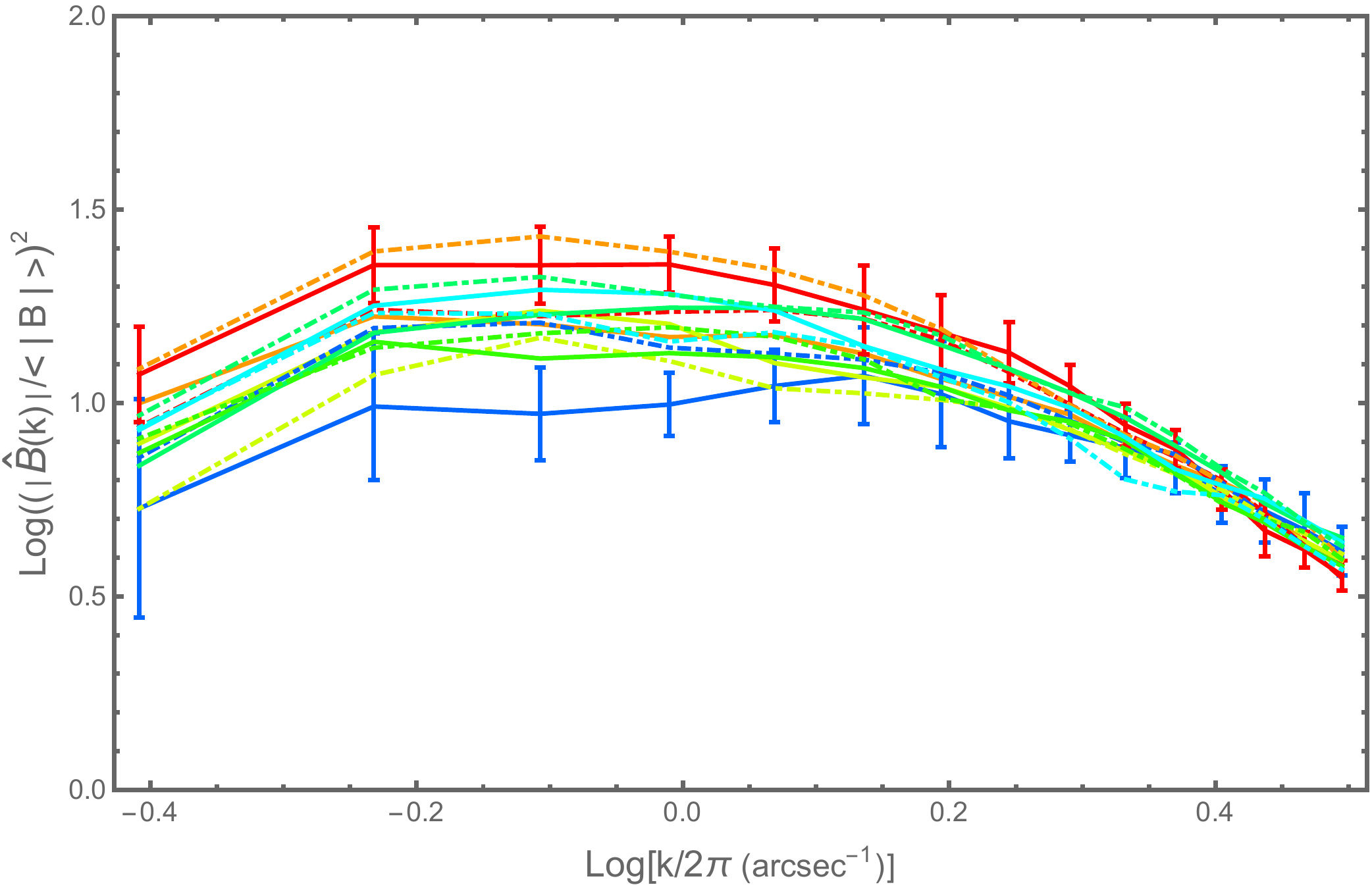}

 \caption{Power spectra of the unsigned magnetic flux density  in the IN regions at high latitudes in both hemispheres. Full lines  refer to the southern hemisphere; dot-dashed lines to the northern hemisphere. The bars show the $\pm\sigma$ intervals for the 2008 and 2014 data. The color code for the year of observation is the same as in Fig. \ref{fig4}.
 }
  \label{fig7}
\end{figure}  
\begin{figure}[ht]
\includegraphics[width=0.45\textwidth]{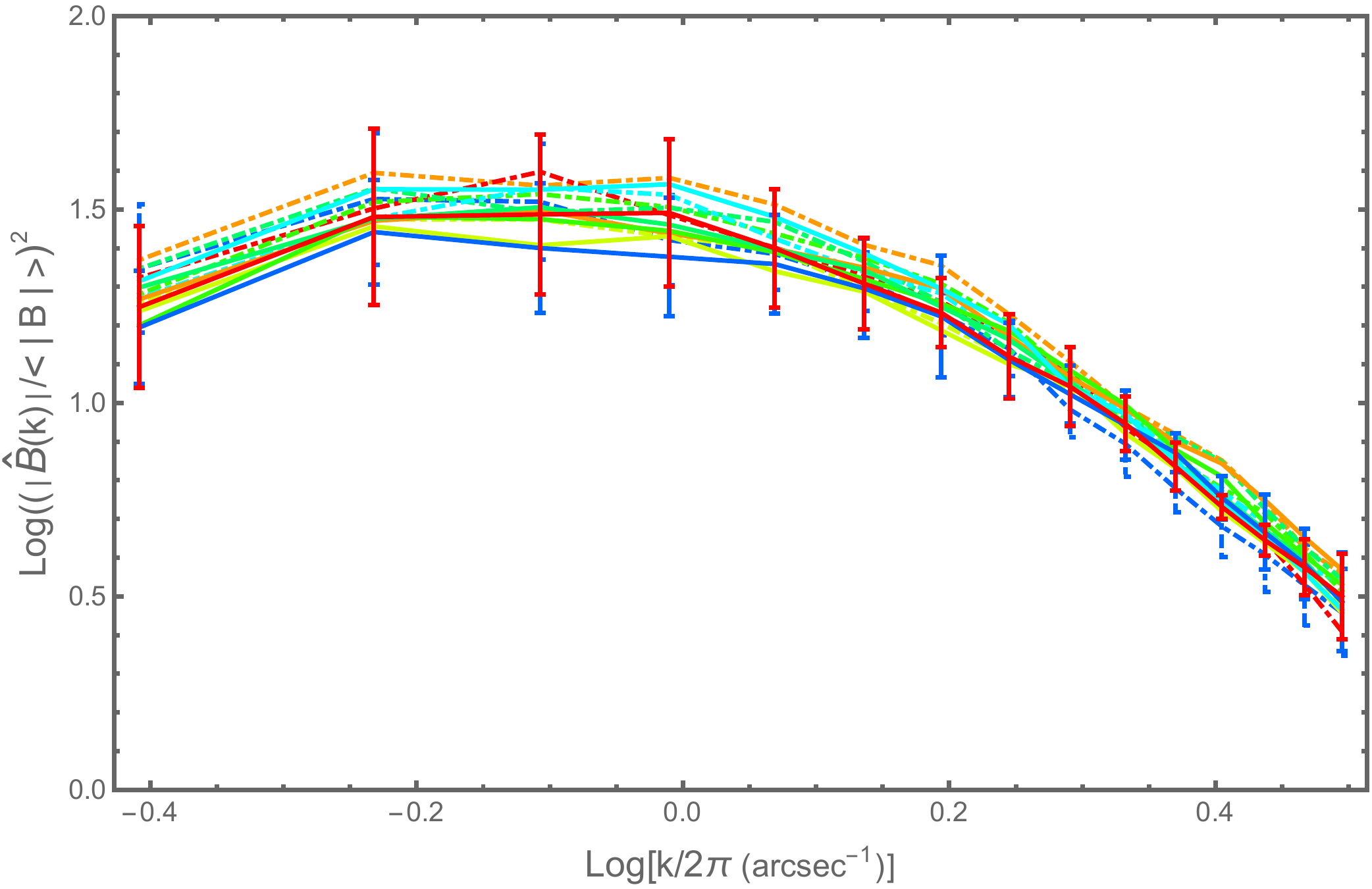}

 \caption{Same as in Fig.\ref{fig7} for IN regions at  
 $\theta= \pm 30 ^\circ$.
 }
  \label{fig8}
\end{figure}

\begin{figure}[ht]
\includegraphics[width=0.45\textwidth]{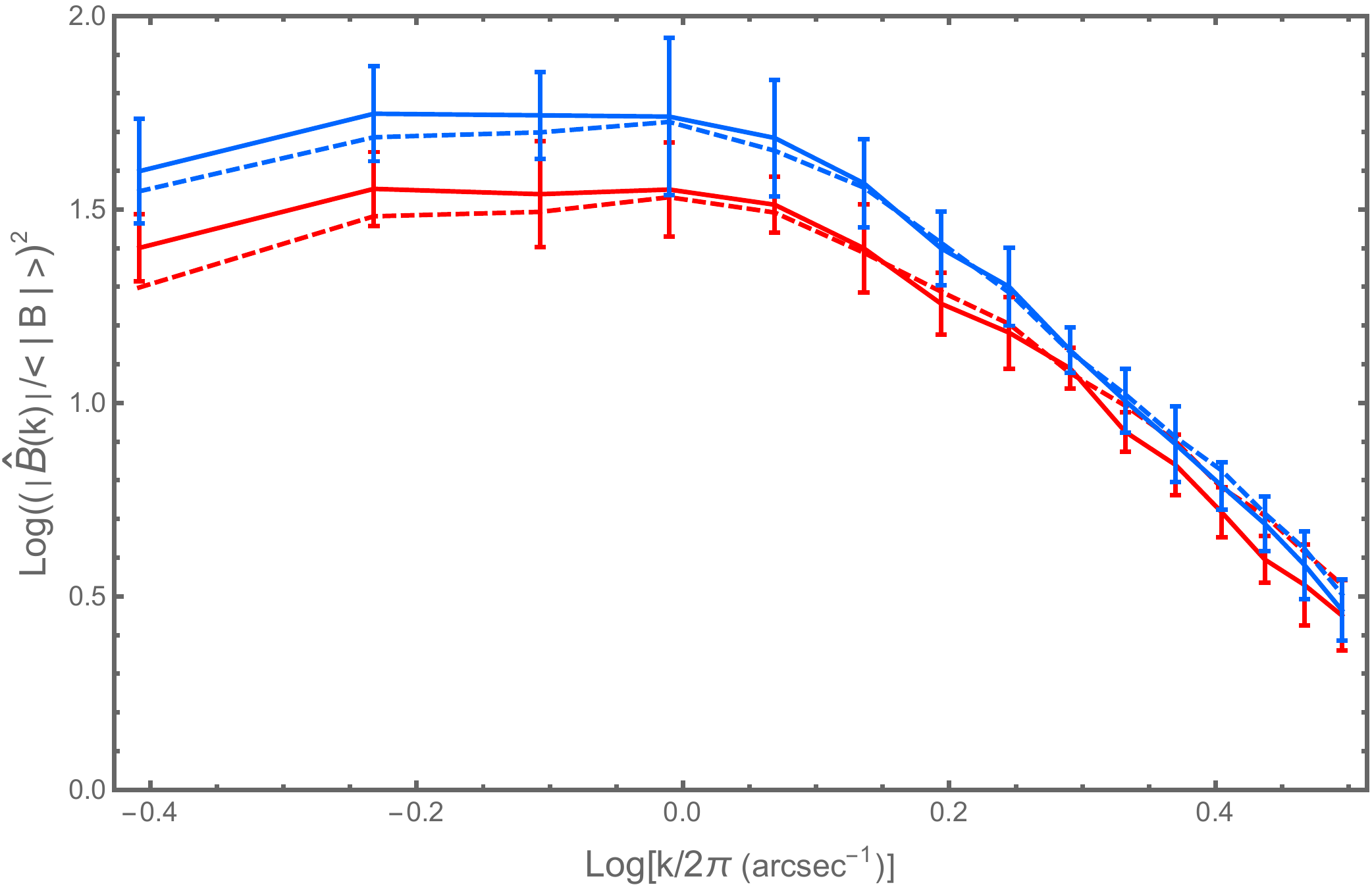}

 \caption{Power spectra of  the  spatial fluctuations of the unsigned magnetic flux density  in the IN regions at the active latitude
 $\theta= - 18 ^\circ$. Full lines: from the FeI 630.25 nm line; dashed lines: from the FeI 630.15 nm line. Blue lines: in 2014; red lines: in 2008. The bars show the $\pm\sigma$ intervals.
 }
  \label{fig10}
\end{figure}

 In Fig.\ref{fig7}  we show the power spectra at high latitudes on both hemispheres from 2008 to 2014. We note that at spatial frequencies higher that  2 arc-second$^{-1}$ (i.e., spatial scales smaller than 0.5''), the  power spectra remain mostly unchanged and  similar on both hemispheres. Figure \ref{fig8} shows the spectra at latitudes of  $\pm 30 ^\circ$, we observe no significant variations with the solar cycle and quite similar spectra on both hemispheres. This is quite typical of most of the results at mid-latitudes, except in some regions in which an enhanced network is visible at solar maximum. This is the case in the 2014 data for the scan at $\theta =-18^\circ$  (the magnetic map is shown in Fig. \ref{fig1} ). Even if we avoided network elements, the mean value of the unsigned longitudinal flux density is significantly higher in the IN regions that we selected from this scan (see Fig. \ref{fig2}). Figure  \ref{fig10} shows the comparison of the reduced power spectra of the magnetic flux fluctuations at solar maximum and minimum at this latitude, we note a marginally significant increase in the power at granular scales at solar maximum but  no variation at scales smaller than 0.5''. 

\section{Discussion and conclusions}
When discussing the results shown above it is important to keep in mind that the longitudinal flux reflects different magnetic 
components at disk center and high latitudes. At disk center the line of sight is vertical so we observe vertical fields, whereas at 
high latitudes it is inclined on the solar surface so we observe mainly the horizontal fields of the selected IN regions.  We recall that IN magnetic fields in the low photosphere have a more uniform distribution of field inclinations than network and other strong-field region and that the inclination increases with height.
As the heliocentric angle is larger at high latitude we also observe higher layers of the photosphere.

Our study shows that vertical fields in the IN at low latitudes do not vary significantly with the solar cycle, at least on the spatial scale that we could investigate, between 0.3 '' and 5 ''.  However we note that in the IN regions located within an enhanced network at solar maximum the mean longitudinal magnetic flux density is increased and the power of the magnetic fluctuations is marginally stronger at granular and larger scales. 

The horizontal fields observed at high latitudes show variations in opposition of phase with the solar cycle, except at  scales smaller than 0.5'' , where the power spectra remain constant.
One straightforward interpretation would be that horizontal  fields at granular and larger scales are affected by the contribution of magnetic elements from decaying active regions migrating toward the poles that are more numerous at solar minimum. 

 At scales smaller than 0.5'' the power spectra of the IN longitudinal magnetic flux density are constant whatever the latitude, indicating the presence of a  time-independent magnetic component. 
 The time-independence of the spectra is a strong indication that the mechanism at the origin of IN fields is not, at least not directly, correlated to the global dynamo. However indirect effects may exist. In the numerical simulations of \citet{Rempel2014, Rempel2018},  where a small-scale dynamo is operating, the mean magnetic strength of the IN fields is increased when the lower boundary condition accounts for a horizontal magnetic field transported from the convective region upward to the photosphere. This could be at the origin of the increase in the mean unsigned longitudinal magnetic flux density and the power of the magnetic fluctuations that we observe in the enhanced network region in our data in 2014. 
%{\bf However an alternative mechanism could be the mixing of active region remnants  by the turbulence in the upper convection zone. It is not clear if such a mixing could operate on the long time-scale }

The power spectra at all latitudes have a broad maximum at sub-granular spatial scales on the order of  900 km (except at  high latitudes at solar maximum). It is difficult to compare our observational results with the numerical simulations of \citet{Rempel2014} because we measured the unsigned longitudinal flux and not the magnetic energy. In the simulations the magnetic energy power spectrum has a broad maximum at small spatial scales on the order of 500 km to 1000 km; this seems consistent with our results. We do not intend to measure the slope index of the power spectra because we think that a meaningful measurement should be done on at least a decade of spatial scales. Furthermore  the limited signal-to-noise ratio of our spectra would not allow a precise determination.  

\citet{Katsukawa2012} also studied the power spectra of the magnetic unsigned vertical and horizontal component at disk center in weak IN regions.  They used a very different method for the deconvolution of the PSF that they applied on the magnetic maps directly, and they obtained spectra with  broad maxima at scales on the order of 800 km. We find a slightly larger scale but as the maximum is very broad a different treatment of the deconvolution may be at the origin of this difference. They did not study the long-term variations of the spectra.

We now discuss the comparison between the magnetic flux density  measured in the two FeI 630 nm lines. We found, quite unexpectedly, that the average value is higher in the maps obtained with the FeI 630.15 nm that is formed higher in the photosphere. This seems to contradict the decrease of the magnetic fields with height that is  derived from the inversion of the Stokes vectors as in \citet{Danilovic2016}. There is no contradiction as far as the power of the spatial fluctuations are concerned; we found lower fluctuations in the maps obtained from the spectral line that is formed higher up in the photosphere. The result about the average unsigned flux density is more puzzling. At high latitudes this can be understood as an increase in the inclination of the field with height that has also been found in recent inversions. At low latitude, where we measure the vertical fields, this could be the result of  more cancellation of flux in unresolved structures at lower depth than at higher altitude. Especially if the IN magnetic field forms loop-like structures at various scales, large-scale loops would reach higher altitudes than the smallest loops.  Therefore the average unsigned flux density over the region could be higher when measured at higher heights where more loop elements are resolved.

%\begin{appendix} 
%\end{appendix}

\begin{acknowledgements}
We thank Bruce Lites for fruitful comments on an earlier version of this paper.
Hinode is a Japanese mission developed and launched by ISAS/JAXA, collaborating with NAOJ as a domestic partner, NASA and STFC (UK) as international partners. Scientific operation of the Hinode mission is conducted by the Hinode science team organized at ISAS/JAXA. This team mainly consists of scientists from institutes in the partner countries. Support for the post-launch operation is provided by JAXA and NAOJ (Japan), STFC (U.K.), NASA, ESA, and NSC (Norway).
\end{acknowledgements}

\bibliography{faurob}

\begin{thebibliography}{15}
\expandafter\ifx\csname natexlab\endcsname\relax\def\natexlab#1{#1}\fi

\bibitem[{{Bellot Rubio} \& {Orozco Su{\'a}rez}(2019)}]{BellotRubio2019}
{Bellot Rubio}, L. \& {Orozco Su{\'a}rez}, D. 2019, Living Reviews in Solar
  Physics, 16, 1

\bibitem[{{Buehler} {et~al.}(2013){Buehler}, {Lagg}, \&
  {Solanki}}]{Buehler2013}
{Buehler}, D., {Lagg}, A., \& {Solanki}, S.~K. 2013, \aap, 555, A33

\bibitem[{{Danilovic} {et~al.}(2016){Danilovic}, {van Noort}, \&
  {Rempel}}]{Danilovic2016}
{Danilovic}, S., {van Noort}, M., \& {Rempel}, M. 2016, \aap, 593, A93

\bibitem[{{Faurobert} \& {Ricort}(2015)}]{Faurobert2015}
{Faurobert}, M. \& {Ricort}, G. 2015, \aap, 582, A95

\bibitem[{{Go{\v s}i{\'c}} {et~al.}(2014){Go{\v s}i{\'c}}, {Bellot Rubio},
  {Orozco Su{\'a}rez}, {Katsukawa}, \& {del Toro Iniesta}}]{Gosic2014}
{Go{\v s}i{\'c}}, M., {Bellot Rubio}, L.~R., {Orozco Su{\'a}rez}, D.,
  {Katsukawa}, Y., \& {del Toro Iniesta}, J.~C. 2014, \apj, 797, 49

\bibitem[{{Jin} {et~al.}(2011){Jin}, {Wang}, {Song}, \& {Zhao}}]{Jin2011}
{Jin}, C.~L., {Wang}, J.~X., {Song}, Q., \& {Zhao}, H. 2011, \apj, 731, 37

\bibitem[{{Katsukawa} \& {Orozco Su{\'a}rez}(2012)}]{Katsukawa2012}
{Katsukawa}, Y. \& {Orozco Su{\'a}rez}, D. 2012, \apj, 758, 139

\bibitem[{{Lites} {et~al.}(2014){Lites}, {Centeno}, \& {McIntosh}}]{Lites2014}
{Lites}, B.~W., {Centeno}, R., \& {McIntosh}, S.~W. 2014, \pasj, 66, 4

\bibitem[{{Lites} \& {Ichimoto}(2013)}]{LitesSP-PREP2013}
{Lites}, B.~W. \& {Ichimoto}, K. 2013, \solphys, 283, 601

\bibitem[{{Quintero Noda} {et~al.}(2015){Quintero Noda}, {Asensio Ramos},
  {Orozco Su{\'a}rez}, \& {Ruiz Cobo}}]{QuinteroNoda2015}
{Quintero Noda}, C., {Asensio Ramos}, A., {Orozco Su{\'a}rez}, D., \& {Ruiz
  Cobo}, B. 2015, \aap, 579, A3

\bibitem[{{Rempel}(2014)}]{Rempel2014}
{Rempel}, M. 2014, \apj, 789, 132

\bibitem[{{Rempel}(2018)}]{Rempel2018}
{Rempel}, M. 2018, \apj, 859, 161

\bibitem[{{Semel}(1970)}]{Semel1970}
{Semel}, M. 1970, \aap, 5, 330

\bibitem[{{Stein} {et~al.}(2011){Stein}, {Lagerfj{\"a}rd}, {Nordlund}, \&
  {Georgobiani}}]{Stein2011}
{Stein}, R.~F., {Lagerfj{\"a}rd}, A., {Nordlund}, {\r{A}}., \& {Georgobiani},
  D. 2011, \solphys, 268, 271

\bibitem[{{Uitenbroek}(2003)}]{Uitenbroek2003}
{Uitenbroek}, H. 2003, \apj, 592, 1225

\end{thebibliography}
\bibliographystyle{aa}

\end{document}